\documentstyle[preprint,aps,psfig]{revtex}
\newcommand{\be}{\begin{eqnarray}}
\newcommand{\e}{\end{eqnarray}}

\begin{document}
\draft
\tighten

\title{Off-forward Matrix Elements in Light-Front Hamiltonian QCD}

\author{A. Mukherjee\thanks{e-mail: mukherje@kph.uni-mainz.de 
(On leave of absence from Saha Institute of Nuclear Physics, Kolkata, India.)}
and M. Vanderhaeghen\thanks{e-mail: marcvdh@kph.uni-mainz.de}\\
 Instit\"ut f\"ur Kernphysik, Universit\"at Mainz, D-55099, Mainz, Germany}

\date{June 18, 2002}

\maketitle
\begin{abstract}
We investigate the off-forward matrix element of the light cone vector 
operator for a dressed quark state in light-front Hamiltonian perturbation 
theory. We obtain the corresponding splitting functions in a straightforward 
way. We show that the end point singularity is canceled by the contribution 
from the normalization of state. Considering  mixing with the gluon operator, 
we verify the helicity sum rule in perturbation theory. We show that the quark
 mass effects are suppressed in the plus component of the matrix element but in
 the transverse component, they are not suppressed. We emphasize that this is 
a particularity of the off-forward matrix element and is absent in the forward case. 
\end{abstract}
\vskip .2in
{\bf 1. Introduction}
\vskip .2in
The hadronic matrix elements of quark and gluon operators appear in the
description of all scattering processes. They are in general of two types:
in the inclusive processes, one encounters diagonal matrix elements of
bilocal operators. These matrix elements are related to parton
distributions. On the other hand, in the elastic exclusive processes, one encounters
form factors, which are off-diagonal matrix elements of local operators. The
 generalized parton distributions interpolate between these two types of
matrix elements \cite{dieter}. These are off-diagonal matrix elements of light-front
bilocal operators. They play an important role in the deeply virtual Compton
scattering amplitude \cite{rad1} and electroproduction of
mesons \cite{rad3,collins} (for reviews of hard exclusive reactions, see
\cite{rev}). The off-forward matrix
elements are the generalizations of the above two types of matrix
elements; 
parton distributions are the forward limits of generalized parton distributions
(GPD) and form factors are moments of them.

Recently, the generalized parton distributions have been investigated in the
light-front formalism by several authors and an overlap representation for
the plus component in
terms of light-front wave functions has been given \cite{brod,kroll}. Also GPD's have been constructed using light cone model wave functions \cite{miller}. The perpendicular and the
minus components are somewhat more complicated because the operators in
these
cases involve interactions. They are the higher twist components.  
In this work, we calculate the the off-forward   
matrix elements of the plus component and the mass dependent helicity-flip part of the 
perpendicular component  of the  bilocal 
vector operator for a dressed quark target in
light-front Hamiltonian QCD. Recently, we have made considerable progress in
understanding polarized and
unpolarized deep inelastic scattering (DIS) structure functions in this approach
\cite{hari} and we have shown that it gives an intuitive picture of DIS. It is suitable 
to calculate the forward
matrix elements of transverse and minus components of the bilocal current
operator. Also, the presence of quark mass does not cause any problem. 
Interference effects are straightforward to handle. The splitting functions
are obtained easily and they agree with the well known expressions
\cite{hari}. It is possible to derive new sum rules which connect DIS structure functions to light-front QCD Poincare generators.
In this work, we extend our approach to off-forward matrix elements.

\vskip .2in
{\bf 2. Plus Component}
\vskip .2in
We work in the so called symmetric frame \cite{brod,kroll}. The momentum of the initial state
is $P^\mu$ and that of the final state is $P'^\mu$. The average momentum
between initial and final state is then ${\bar P}^\mu={P^\mu+P'^\mu\over 2}$.

The momentum transfer is given by $\Delta^\mu=P'^\mu-P^\mu$, $ P'_\perp =
-P_\perp
={\Delta_\perp\over 2}$, skewedness $\xi=-{\Delta^+\over {2 {\bar P}^+}}$.
Without any loss of generality, we take $\xi >0$. We also get $\Delta^-={\xi
{\bar P}^2\over {\bar P}^+}$.

The off-forward matrix element  is given by,
\be
F^\mu_{\lambda \lambda'}=\int {dz^-\over {2 \pi}} e^{{i\over 2}{\bar x}{\bar P}^+
{z^-}}\langle P'
\lambda' \mid \bar \psi (-{ z^-\over 2}) \gamma^\mu\psi ({ z^-\over 2}) 
\mid P \lambda \rangle.
\label{eq1}
\e
\vspace*{-0.4cm}
\begin{center}
\psfig{figure=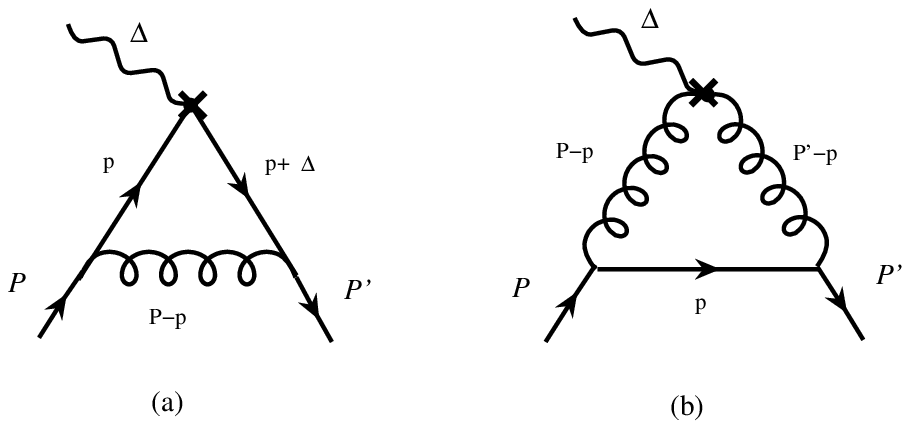,width=18.0cm,height=7.0cm}
\end{center}
\vspace*{-0.4cm}
\begin{center}
\parbox{14.0cm}
{{\footnotesize
Fig. 1: Light-front time ordered diagrams considered in the calculations of 
Eqs. (\ref{eq1}) and (\ref{gg}); only one time ordering is important in the kinematical region considered.}}
\end{center}
\vspace{0.2cm}
The $+$ component of the matrix element is parametrized in terms of
the off-forward distributions $H({\bar x},\xi,t)$ and $E({\bar x},\xi,t)$.
However, the matrix element can be expressed directly in terms of overlaps
of light-front wave functions. We calculate the plus component of this matrix 
element taking  the target to be a dressed quark to order $g^2$ (see Fig.
1 (a)). 
We work in the light-front gauge,
$A^+=0$, where the path-ordered exponential between the fermion fields in
the bilocal operator is unity. For simplicity we suppress the flavor indices.   

The Fock space expansion of the operator is given by,
\be
O^+&=& 4 \sum_s\int {dk^+ d^2k^\perp\over {2(2 \pi)^3 \sqrt k^+}}\int {dk'^+ 
d^2k'^\perp\over {2(2 \pi)^3 \sqrt k'^+}} \nonumber\\&&~~~~
\Big [ \delta(2 {\bar x} {\bar
P}^+-k'^+-k^+) b^\dagger (k,s)b(k',s)\nonumber\\&&~~~~~~~~~~~~~~~
+ \delta(2 {\bar x} {\bar
P}^++k'^++k^+) d (k,-s)d^\dagger (k',-s)
\nonumber\\&&~~ + \delta(2 {\bar x} {\bar P}^++k^+-k'^+) d (k,-s) b(k',s)
\nonumber\\&&~~~~~~~~~~~~~~~~ 
+ \delta(2 {\bar x} {\bar P}^++k'^+-k^+) b^\dagger(k,s)
d^\dagger(k',-s)\Big ].
\label{fock}
\e
We have, $k^+>0$,$k'^+>0$, $k^+-k'^+=p^+-p'^+=2 \xi {\bar p}^+$. In the
kinematical region, $\xi<{\bar x}<1$, only the first term in Eq.
(\ref{fock}) contributes
\cite{kroll}. We
restrict ourselves to this kinematical region. 

We take the state 
$ \mid P, \sigma \rangle$ to be a dressed quark
consisting of bare states of a quark and a quark plus 
a gluon:
\begin{eqnarray}
\mid P, \sigma \rangle && = \phi_1 b^\dagger(P,\sigma) \mid 0 \rangle
\nonumber \\  
&& + \sum_{\sigma_1,\lambda_2} \int 
{dk_1^+ d^2k_1^\perp \over \sqrt{2 (2 \pi)^3 k_1^+}}  
\int 
{dk_2^+ d^2k_2^\perp \over \sqrt{2 (2 \pi)^3 k_2^+}}  
\sqrt{2 (2 \pi)^3 P^+} \delta^3(P-k_1-k_2) \nonumber \\
&& ~~~~~\phi_2(P,\sigma \mid k_1, \sigma_1; k_2 , \lambda_2) b^\dagger(k_1,
\sigma_1) a^\dagger(k_2, \lambda_2) \mid 0 \rangle. 
\label{eq2}
\end{eqnarray} 
Here $a^\dagger$ and $b^\dagger$ are bare gluon and quark
creation operators respectively and $\phi_1$ and $\phi_2$ are the
multiparton wave functions. They are the probability amplitudes to find one
bare quark and one quark plus gluon inside the dressed quark state
respectively. Up to one loop, if one considers all kinematical
regions, there will be non-vanishing contributions from the overlap of
3-particle and one particle sectors of the state, this situation is similar
to QED \cite{brod}. In the kinematical region 
we are considering, such kind of overlaps are absent and it is sufficient
to consider dressing only by a single gluon. The state is normalized to one.
$\phi_1$ actually gives the normalization constant of the state \cite{hari}:
\be
{\mid \phi_1 \mid}^2=1-{\alpha_s\over {2 \pi}} C_f
\int_\epsilon^{1-\epsilon} dx {{1+x^2}\over {1-x}}log
{Q^2\over \mu^2},
\label{c5nq}
\e
within order $\alpha_s$.  Here $\epsilon$ is a small cutoff on $x$.
 
The matrix element becomes,
\be
F^+&=& 4{\sqrt
P^+}{\sqrt P'^+}\Big [ \phi_1^*(P') \phi_1(P) \delta(2 {\bar x}
{\bar P}^+-2{\bar P}^+)\nonumber\\&& + \sum \int dp_1^+d^2p_1^\perp
\phi_2^*(P',\Delta+p_1,P-p_1)
\phi_2(P,p_1,P-p_1) \delta(2 {\bar x}{\bar P}^+-\Delta^+-2 p_1^+)\Big ]
\e
$ \sum $ denotes summation over helicities of the quark and gluon.
The first term comes from one particle sector and it contributes only when
${\bar x}=1$. The first term receives contribution upto order $\alpha_s$
from the normalization condition of the state.

The other nontrivial contribution comes from the two particle
sector given by the second term.

We introduce Jacobi momenta $x_i$,${q_i}^\perp$ such that $\sum_i x_i=1$ and
$\sum_i {q_i}^\perp=0$. Also, we introduce boost invariant wave functions,
\be
\psi_1=\phi_1, ~~~~~~~~~~~\psi_2(x_i,q_i^\perp)= {\sqrt P^+} \phi
(k_i^+,{k_i}^\perp).
\e
The contribution from the two-particle sector then becomes, 
\be
F^+= 2\sum \int d^2q^\perp \psi^*_2({{\bar x}-\xi\over {1-\xi}},q^\perp+{1-{\bar
x}\over {1-\xi^2}}\Delta^\perp) \psi_2({{\bar x}+\xi\over {1+\xi}}, q^\perp).
\label{matrix+}
\e
The two particle wave function depends on the helicities of the quark and gluon and is given in terms of $\psi_1$ as,
\be
\psi^\sigma_{2\sigma_1,\lambda}(x,q^\perp)&=& -{x(1-x)\over (q^\perp)^2}T^a 
{1\over {\sqrt {(1-x)}}} {g\over
{\sqrt {2(2\pi)^3}}} \chi^\dagger_{\sigma_1}\Big [ 2 {q^\perp\over
{1-x}}+{{\tilde \sigma^\perp}\cdot q^\perp\over x} {\tilde \sigma^\perp}
\nonumber\\&&~~~~~~~~~~~~~~~~~~
-i m{\tilde \sigma}^\perp {(1-x)\over x}\Big ]\chi_\sigma \epsilon^{\perp *}_\lambda \psi_1.
\label{psi2}
\e
Here ${\tilde \sigma}^1=\sigma^2$ and  ${\tilde \sigma}^2=-\sigma^1$.
Using Eq. (\ref{matrix+}) and Eq. (\ref{psi2}), we see that the mass terms give suppressed contributions. 
Since mass terms  in the vertex cause helicity flip, helicity flip parts of 
the matrix element are suppressed. We calculate the helicity non-flip part.

Using Eq. (\ref{psi2}) we get, from Eq. (\ref{matrix+}),
\be
F^+=  \int d^2q^\perp {g^2\over {(2 \pi)^3}} C_f{(1-\xi^2)^{1\over 2}\over (1-{\bar
x})} {q^\perp \cdot (q^\perp+{(1-{\bar x})\over (1-\xi^2)}\Delta^\perp)\over
{(q^\perp)^2(q^\perp +{(1-{\bar x})\over (1-\xi^2)}\Delta^\perp)^2}}(1-2
\xi^2+{\bar x^2}),
\e
where $C_f={(N^2-1)\over 2N}$ for $SU(N)$.
The $q^\perp$ integral is nontrivial and it is divergent for large $q^\perp$.
Integration over the polar angle gives,
\be
\int d^2q^\perp {q^\perp (q^\perp+{(1-{\bar x})\over (1-\xi^2)}\Delta^\perp)
\over
{(q^\perp)^2(q^\perp+a \Delta^\perp)^2}} &=&2 \pi
\int_\mu^\Lambda q dq {1\over {\sqrt {(q^2+a^2 \Delta^2)^2-4a^2\Delta^2q^2}}
}\nonumber\\&&~~~-
\pi \int_\mu^\Lambda {dq\over q} \Big [{(q^2+a^2 \Delta^2)\over 
{\sqrt {(q^2+a^2\Delta^2)^2-4a^2\Delta^2q^2}}}-1 \Big ]\nonumber\\&&
=I_1+I_2,
\e  
where {$a={(1-{\bar x})\over (1-\xi^2)}$. $\Lambda$ is the large transverse
momentum cutoff and $\mu$ is the factorization scale separating hard and
soft dynamics \cite{hari}. The divergence structure of the above integral can be seen by
expanding the denominator of the integrand in the limit of small
$\Delta=\mid \Delta^\perp \mid$.
We find that $I_1$ is logarithmically divergent in the limit $\Lambda
\rightarrow \infty$ but $I_2$ has no divergent part. 
The full divergent part is given by,
\be
F^+= 2 {g^2\over (2 \pi)^3}C_f{{\sqrt {1-\xi^2}}\over (1-{\bar x})}{(1-2
\xi^2+{\bar x}^2)\over (1-\xi^2)} \pi log
{Q^2\over \mu^2},
\e
where we have cut the transverse momentum integral at some scale $Q^2$.
The splitting function can be easily extracted from the above expression :
\be
P_{qq}({\bar x},\xi)=C_f {(1+{\bar x}^2-2\xi^2)\over {(1-{\bar x})
(1-\xi^2)}}.
\e
This agrees with \cite{ji} (with ${\xi\over 2}$ replaced by $\xi$). 
The above expression contains end point
singularity at ${\bar x}=1$. However, this is canceled by the contribution from 
the normalization of the state to the single particle matrix element,
 so that the final result becomes \footnote{Here ${1\over (1-x)_+}$ is the usual (principal value) 
plus prescription.},
\be
F^+=2{\sqrt {1-\xi^2}} \Big [ \delta(1-{\bar x})+{\alpha_s\over {2
\pi}}C_f log{Q^2\over \mu^2}\Big ( {3\over 2} \delta(1-{\bar x}) +
{(1+{\bar x}^2-2\xi^2)\over {(1-{\bar x})_+(1-\xi^2)}}\Big ) \Big ].
\label{plusres}
\e 

This result shows the importance of the normalization contribution to the
single particle matrix element. In other words, it includes contributions
from virtual gluon emission.
 In this approach, one uses probability
amplitudes rather than probability densities in Altarelli-Parisi method 
and the effects due to both
real and virtual gluon emissions are taken into account to the same order in
$\alpha_s$ without any difficulty.  In order to get the
full scale evolution, one has to consider all the 
kinematical sectors which is beyond the scope of the present work. Also, one can see
that the final result has no singularity at ${\bar x}=\xi$ which is as
expected \cite{brod}.

Next, we calculate the helicity flip part of the matrix element. The
helicity flip contributions come from the mass term in the expression of the
two-particle wave function. The form of the wave function  shows that this
contribution is suppressed.

Next, we parametrize the off-forward matrix element in terms of the generalized quark
distributions,
\be
F^\mu_{\lambda \lambda'} = {1\over { {\bar P}^+}} 
{\bar U}_{\lambda'}(P') 
\Big [ H_q({\bar x},\xi,t) \gamma^\mu + E_q({\bar x},\xi,t) {i\over {2M}} 
\sigma^{\mu \alpha} \Delta_\alpha \Big ] U_\lambda(P) +.....
\label{para}
\e
where the ellipses indicate higher twist terms.
$U_\lambda(P)$ is the quark spinor in our case. Using the explicit form of the 
 light front spinors, we get for the plus component,
\be
F^+\delta_{\lambda' \lambda}=2 {\sqrt {1-\xi^2}} H_q({\bar x},\xi,t)-{2 \xi^2 \over {\sqrt {1-\xi^2}}}
E_q({\bar x},\xi,t).
\label{he}
\e
Here we have calculated the helicity non-flip part.

The helicity flip part becomes,
\be
F^+ \delta_{\lambda' -\lambda}= {-\Delta^1+i\Delta^2\over {m \sqrt
{1-\xi^2}}}E_q({\bar x},\xi,t).
\e
Since the helicity flip part of the matrix element is suppressed, we
naturally find that 
$E$ is suppressed in perturbation theory. Here we have taken both $m$ and
 $\mid \Delta^\perp \mid$ to be small. In the limit of small $\xi$, we expand $H_q({\bar x},\xi,t)\approx
H_q({\bar x},0,0)+\xi^2 H_q'({\bar x},0,0)$ in the kinematical region
 $\xi < {\bar x}<1$.
Using Eq. (\ref{plusres}) and Eq. (\ref{he}) and equating the coefficients of equal powers of 
$ {\xi} $ on both sides, we get,
\be
H_q({\bar x},0,0)=\delta(1-{\bar x})+{\alpha_s\over {2
\pi}}C_f log{Q^2\over \mu^2}\Big ( {3\over 2} \delta(1-{\bar x}) +
{(1+{\bar x}^2)\over {(1-{\bar x})_+}}\Big ) 
\label{hh}
\e
\be
H_q'({\bar x},0,0)=-{\alpha_s\over {2
\pi}}C_f log{Q^2\over \mu^2}(1+{\bar x}).
\label{ee}
\e
 Eq. (\ref{hh}) gives the forward limit of the generalized quark distribution 
$H_q$ and it 
can be easily identified with the 
unpolarized quark distribution for a dressed quark state in perturbation
theory \cite{hari}. 
\vskip .2in 
Subsequently, we calculate the gluon distribution,
\be
F_{g \lambda' \lambda}^+=-{1\over {{\bar x} {\bar P}^+}}\int {dz^-\over {2 \pi}} e^{{i\over 2}{\bar P}^+
{ z^-} {\bar x}}\langle P'
\lambda' \mid F^{+ \alpha}(-{ z^-\over 2}) F_\alpha^+ ({ z^-\over 2}) 
\mid P \lambda \rangle.
\label{gg}
\e 
The Fock space expansion of the relevant part of the operator is given by,
\be
O_g={2\over {{\bar x}{\bar P}^+}} {1\over (2(2 \pi)^3)^2}\sum_\lambda \int dk_1^+ d^2k_1^\perp \int dk_2^+ d^2k_2^\perp a^\dagger(k_1,\lambda) a(k_2,\lambda) \delta(2 {\bar x}{\bar P}^+-k_1^+-k_2^+)
\e
We calculate the matrix element for a quark state dressed with a gluon (see
Fig. 1(b)). The
Fock space expansion of the state is given by Eq. (\ref{eq2}).

The matrix element is given by,
\be 
F_g^+ = {2\over {\bar x}}\sum \int d^2q^\perp \psi_2^*({1-{\bar x}\over 
{1-\xi}},q^\perp) \psi_2({1-{\bar x}\over {1+\xi}}, q^\perp+{1-{\bar x}\over 
(1-\xi^2)}\Delta^\perp) \sqrt {{\bar x}^2-\xi^2}.
\label{matrixg}
\e
Using the full form of the two particle wave function, we find that the helicity flip terms proportional to the quark mass give suppressed contribution and the helicity non-flip part is given by,
\be
F_g^+=2 \sqrt {1-\xi^2} {\alpha_s\over 2 \pi}C_f log{Q^2\over \mu^2} {(1+(1-{\bar x})^2-\xi^2)\over {x (1-\xi^2)}}.
\e
The splitting function can be easily extracted and agrees with \cite{ji} with ${\xi\over 2}$ replaced by $\xi$. 

The gluon matrix element is parametrized in terms of the twist two distributions, $H_g$ and $E_g$,
\be
 F^\mu_{g\lambda' \lambda} = {1\over { {\bar P}^+}} 
{\bar U}_{\lambda'}(P') 
\Big [ H_g({\bar x},\xi,t) \gamma^\mu + E_g({\bar x},\xi,t) {i\over {2M}} 
\sigma^{\mu \alpha} \Delta_\alpha \Big ] U_\lambda(P) +.....
\label{parag}
\e
The fact that the helicity flip part of the matrix element is suppressed 
means that  
$E_g$ is also suppressed. In the limit of small $\xi$, we again expand 
 $H_g({\bar x},\xi,t)\approx
H_g({\bar x},0,0)+\xi^2 H_g'({\bar x},0,0)$  in the kinematical region  $\xi < {\bar x}<1$ and we obtain,
\be
H_g({\bar x},0,0)= {\alpha_s\over 2 \pi}C_f log{Q^2\over \mu^2} {(1+(1-{\bar x})^2)\over {\bar x}},
\e
One can identify the above expression with the unpolarized gluon distribution for a dressed quark target \cite{hari}.
\be
 H_g'({\bar x},0,0)= {\alpha_s\over 2 \pi}C_f log{Q^2\over \mu^2}{(1-{\bar x})^2\over {\bar x}}.
\e

We next verify the helicity sum rule in perturbation theory. 
It has been shown \cite{oam} that the light front helicity operator when 
expressed entirely in terms of the dynamical fields in the light front gauge, 
has the same form as in the free theory, provided we restrict ourselves to the
 topologically trivial sector (i. e.  we take the dynamical fields to vanish at the boundary).  This eliminates the residual gauge degrees of freedom and removes the surface terms. The helicity operator is given by,
\be
J^3=J^3_{fi}+J^3_{fo}+J^3_{gi}+J^3_{go},
\e
where, $J^3_{fi}$ is the intrinsic quark helicity, $J^3_{fo}$ is the orbital quark helicity, $J^3_{gi}$ is the intrinsic gluon helicity and 
$J^3_{go}$ is the orbital gluon helicity. The operators are given by,
\be
J^3_{fo}=\int dx^- d^2x^\perp \psi^{+\dagger} i(x^1 \partial^2-x^2 \partial^1) \psi^+,
\e
\be
J^3_{fi}={1\over 2}\int dx^- d^2x^\perp \psi^{+\dagger} \Sigma^3 \psi^+,
\e
\be
J^3_{go}={1\over 2}\int dx^- d^2x^\perp [x^1 (\partial^+A^1 \partial^2 A^2 +  
\partial^+A^2 \partial^2 A^2)-x^2 (\partial^+A^1 \partial^1 A^1 +  
\partial^+A^2 \partial^1 A^2)],
\e
\be
J^3_{gi}= {1\over 2}\int dx^- d^2x^\perp(A^1 \partial^+ A^2-A^2 \partial^+ A^1).
\e
The color indices are implicit in the above expressions.
We can check explicitly using the above expressions that  
the helicity sum rule for a dressed quark in perturbation theory is given 
entirely in terms of $H({\bar x},0,0)$,
\be
\int_0^1 d{\bar x} {\bar x} H_q({\bar x},0,0)=1-{\alpha_s\over {2 \pi}}
C_f log{Q^2\over \mu^2}{4\over 3}= {2\over N}\langle P, \uparrow\mid J^3_{fi}\mid P,\uparrow \rangle+{2\over N}\langle P, \uparrow\mid J^3_{fo}\mid P,\uparrow \rangle,
\label{jf}
\e 
\be
\int_0^1 d{\bar x} {{\bar x} H_g(\bar x},0,0)={\alpha_s\over {2 \pi}}C_f log{Q^2\over \mu^2}{4\over 3}= {2\over N}\langle P, \uparrow\mid J^3_{gi}\mid P,\uparrow \rangle+{2\over N}\langle P, \uparrow\mid J^3_{go}\mid P,\uparrow \rangle,
\label{jg}
\e 
where $N$ is a normalization constant. This gives, for a dressed quark state in perturbation theory, 
\be
\int_0^1 d{\bar x} {\bar x} (H_q({\bar x},0,0)+   H_g({\bar x},0,0))={1\over N}\langle P, \uparrow\mid 2 J^3\mid P,\uparrow \rangle=1.
\e
By comparing the {\it lhs} and {\it rhs} of Eqs. (\ref{jf}) and (\ref{jg}), one 
also verifies  that in perturbation theory the total quark (gluon) momentum contribution is identical to the total quark (gluon) helicity contribution. This result  is expected from the fixed point solutions of the leading log evolution equations of $J_q$ and $J_g$ calculated in \cite{equal}, which shows that the partition of the nucleon spin between quarks and gluons follows the  partition of
 nucleon momentum.
\vskip .2in 
{\bf 3. Helicity Flip Part of the Transverse Component}
\vskip .2in

Having studied the {\it plus} component of the matrix element of the bilocal operator, we now show the importance of quark mass in the matrix element of the 
{\it transverse} component of the bilocal vector operator.
The matrix element of the
transverse component is given by:
\be
F^\perp_{\lambda' \lambda}=\int {dz^-\over {2 \pi}} e^{{i\over 2}
{\bar P}^+{ z}^- {\bar x}}\langle P'
\lambda' \mid \bar \psi (-{z^-\over 2}) \gamma^\perp \psi ({z^-\over 2}) 
\mid P \lambda \rangle.
\label{eqt}
\e
The bilocal operator in this case can be written as,
\be
O=\bar \psi (-{z^-\over 2}) \gamma^\perp \psi ({z^-\over 2})
= \psi^{+ \dagger}(-{z^-\over 2}) \alpha^\perp \psi^{- }({z^-\over 2})
+\psi^{-\dagger}(-{z^-\over 2}) \alpha^\perp \psi^{+ }({z^-\over 2}).
\label{hc}
\e
The operator involves the constrained field $\psi^{- }({ z^-\over 2})$ and
therefore it is the so-called bad component. $\psi^{-}$ can be eliminated
using the constraint equation  in light-front gauge,
\be
i \partial^+ \psi^- = \big [ \alpha^\perp \cdot ( i \partial^\perp + g
A^\perp) + \gamma^0 m \big ] \psi^+.  
\e
The terms linear in mass produce helicity-flip contributions in the matrix element. Quadratic mass terms generate helicity non-flip terms but it can be shown that they are suppressed.  

It is known that in the forward case, for non-zero $P^\perp$, the above
matrix element is related to the unpolarized  DIS structure function $F_2$, 
as calculated through 
the plus component \cite{matrix}. The contribution of the mass dependent helicity-flip
part of the operator in this case cancels between the two terms of Eq.
(\ref{hc}), as a result, quark mass  effects are still suppressed  here in the
forward limit.

We calculate the matrix element of the helicity flip part of the operator 
in the off-forward case.

The operator is given by,
\be
O=O_m+O_{k^\perp}+O_g,
\e
where $O_m$ is the explicit mass dependent part of the operator, $O_{k^\perp}$ is the explicit $k_\perp$ dependent part and $O_g$ gives the interaction dependent part. The Fock space expansion of $O_m$ is given by,
\be
O_m &=&-\psi^{+\dagger}(-{{ z^-}\over 2}) \gamma^\perp {m\over i\partial^+}
\psi^+({{z^-}\over 2})+[-{m\over i\partial^+}\psi^{-\dagger}(-{{ 
z^-}\over 2})] \gamma^\perp \psi^+({{z^-}\over 2})
\nonumber\\&&~~~~~
= 2 \sum \int {dk^+ d^2k^\perp\over {2(2 \pi)^3 \sqrt k^+}}\int {dk'^+ 
d^2k'^\perp\over {2(2 \pi)^3 \sqrt k'^+}} \nonumber\\&&~
(im) \delta(2 {\bar x} {\bar P}^+-k^+-k'^+) b^\dagger(k,s)b(k',s')
\chi^\dagger_s\sigma^2 \chi_{s'} ({1\over k^+}-{1\over k'^+}).
\e
Here $\chi$ is the two component spinor.
This is the part of the operator that is relevant in the kinematical region 
$\xi<{\bar x}<1$. 
We take the state to be a  quark dressed with a gluon as before. The matrix element is given by,
\be
F^\perp_m &=& -{2 \xi\over {\bar P}^+} (im) \chi^\dagger_\sigma \sigma^2
 \chi_{\sigma'} {1\over \sqrt {1-\xi^2}} \Big [ \delta(1-{\bar x})\psi_1^*\psi_1\nonumber\\&&+\int d^2
q^\perp {1\over {{\bar x}^2-\xi^2}} \psi_2^*({{\bar x}-\xi\over {1-\xi}},
q^\perp+{1-{\bar
x}\over {1-\xi^2}}\Delta^\perp) \psi_2({{\bar x}+\xi\over {1+\xi}}, q^\perp)\Big ].
\e
Using the explicit form of the two particle wave function, and also using the normalization condition of the state, we write this as,
\be
F^\perp_m &=& -{2 \xi\over {\bar P}^+} (im) \chi^\dagger_\sigma \sigma^2
 \chi_{\sigma'}{1\over \sqrt {1-\xi^2}}\Big [ \delta(1-{\bar x})\nonumber\\&&
+{\alpha_s\over {2
\pi}}C_f log{Q^2\over \mu^2}\Big ( {3\over 2} \delta(1-{\bar x}) +
{2{\bar x}-2\xi^2\over {(1-{\bar x})_+({\bar x}^2-\xi^2)}}\Big ) \Big ].
\label{mm}
\e
The other contribution to the helicity flip matrix element comes from,
\be
O_{k^\perp}=\xi^\dagger(-{{z}^-\over 2})(\partial^1+i\sigma_3\partial^2)
{1\over \partial^+} \xi({{z}^-\over 2})+
\Big [(\partial^1-i\sigma_3\partial^2){1\over \partial^+}\xi^\dagger
(-{{z}^-\over 2})\Big ] \xi({{z}^-\over 2}).
\e
Here $\xi$ is the two component fermion field.
The Fock space expansion is given by,
\be
F^1_{k_\perp} &=&  \sum \int {dk^+ d^2k^\perp\over {2(2 \pi)^3 \sqrt k^+}}\int {dk'^+ 
d^2k'^\perp\over {2(2 \pi)^3 \sqrt k'^+}} \nonumber\\&&~~~~
\Big [ \delta(2 {\bar x} {\bar
P}^+-k'^+-k^+) b^\dagger (k,s)b(k',s) \Big [ {k^j\over k^+}
(\delta^{ij} -i \sigma_3 \epsilon^{ij})+{k'^j\over k'^+}
(\delta^{ij} +i \sigma_3 \epsilon^{ij}) \Big ].
\e
The two particle contribution to the matrix element is of the form,
\be
F^1_{k^\perp}&=& {1\over {\bar P}^+}\sum \int d^2 q^\perp \psi_2^*({{\bar x}-
\xi\over {1-\xi}},
q^\perp+{1-{\bar
x}\over {1-\xi^2}}\Delta^\perp) \psi_2({{\bar x}+\xi\over {1+\xi}}, q^\perp) 
q^1{ 2{\bar x}\over {{\bar x}^2-\xi^2}}+\nonumber\\&&~~~~+
 {1\over {\bar P}^+}\sum \int d^2 q^\perp \psi_2^*({{\bar x}-\xi\over {1-\xi}},
q^\perp+{1-{\bar
x}\over {1-\xi^2}}\Delta^\perp) \psi_2({{\bar x}+\xi\over {1+\xi}}, q^\perp) 
\chi^{\dagger} (-i \sigma^3 q^2 ) \chi    {2 \xi  \over {{\bar x}^2-\xi^2}}.
\e
                                                              We have taken $\mid \Delta^\perp \mid$ to be small.

The terms linear in mass in $\psi_2$ give the helicity flip contribution given by,
\be
F^1_{k^\perp}={2 \xi\over {\bar P}^+} (im) \chi^\dagger_\sigma \sigma^2
 \chi_{\sigma'} {1\over \sqrt {1-\xi^2}} C_f log{Q^2\over \mu^2}{\alpha_s\over 
{2 \pi}}{2{(1-\bar x}) ({\bar x}^2+\xi^2+2 {\bar x})\over {({\bar x}^2-\xi^2)(1-\xi^2)}}.
\label{mk}
\e
The interaction part of the operator is given by,
\be
F^\perp_g=g \xi^\dagger(-{{z}^-\over 2}) {1\over i\partial^+}(A^1+i\sigma_3 A^2)\xi({{\bar z}^-\over 2})+g \Big [ {1\over {-i \partial^+}} 
 \xi^\dagger(-{{z}^-\over 2})(A^1-i\sigma_3 A^2)\Big ]
\xi({{z}^-\over 2}).
\e
The Fock space expansion of the operator:
\be
F^1_g &=& \sum_{s_1,s_2,\lambda} \int {dk_1^+ d^2 k_1^\perp\over {2(2 \pi)^3 \sqrt k_1^+}}
\int {dk_2^+ 
d^2 k_2^\perp\over {2(2 \pi)^3 \sqrt k_2^+}}\int {dk_3^+ d^2 k_3^\perp
\over {2(2 \pi)^3 
 k_3^+}}\Big (b^\dagger
(k_1,s_1)b(k_2,s_2)a(k_3,\lambda)\nonumber\\&&~~~~~~~~\Big [
4 \delta(2 {\bar x} {\bar
P}^+-k_1^+-k_2^++k_3^+) 
{\chi^\dagger_{s_1}(\epsilon^1 -i \sigma_3 \epsilon^2)\chi_{s_2}\over
{k_1^+-k_3^+}}\nonumber\\&&~~~~~~~~~~~~~~~~~+
4 \delta(2 {\bar x} {\bar
P}^+-k_1^+-k_2^+-k_3^+)
{\chi^\dagger_{s_1}(\epsilon^1 +i \sigma_3 \epsilon^2)\chi_{s_2}  \over
{k_2^++k_3^+}}\Big ]\nonumber\\~~~~~~~~~~&&+ b^\dagger
(k_1,s_1)b(k_2,s_2)a^\dagger(k_3,\lambda)
\Big [4 \delta(2 {\bar x} {\bar
P}^+-k_1^+-k_2^+-k_3^+){\chi^\dagger_{s_1}
(\epsilon^1 -i \sigma_3 \epsilon^2)\chi_{s_2} \over
{k_1^++k_3^+}}\nonumber\\&&~~~~~~~~~~~~~~~~~~+
4 \delta(2 {\bar x} {\bar
P}^+-k_1^+-k_2^++k_3^+~){\chi^\dagger_{s_1}
(\epsilon^1 +i \sigma_3 \epsilon^2) \chi_{s_2} \over {k_2^+-k_3^+}}\Big ]\Big ).
\e
The terms containing $d$ and $d^\dagger$ will contribute in higher order. Out of the four terms, only the first and the fourth terms can contribute to the matrix element. These contributions are given in terms of the overlap of two-particle and one-particle wave functions. However, these contributions are zero
 since $\sum_\lambda\chi^\dagger_{\sigma}(\epsilon^1_\lambda 1 -i \sigma_3 
\epsilon^2_\lambda)
({\tilde \sigma}^\perp \cdot \epsilon^{\perp *}_\lambda))\chi_{\sigma'}=0$.

So we get, from Eq. (\ref{mm}) and (\ref{mk}), the helicity flip part of the matrix element:
\be
F^1&=& -{2 \xi\over {\bar P}^+} (im) \chi^\dagger_\sigma \sigma^2
 \chi_{\sigma'} {1\over \sqrt {1-\xi^2}}\Big [ \delta(1-{\bar x})\nonumber\\&&
+{\alpha_s\over {2
\pi}}C_f log{Q^2\over \mu^2}\Big ( {3\over 2} \delta(1-{\bar x}) +
{2{\bar x}-2\xi^2\over {(1-{\bar x})_+({\bar x}^2-\xi^2)}}\Big ) \nonumber\\&&~~~~~~~-{2\alpha_s\over {2\pi}}C_f log{Q^2\over \mu^2}
{({\bar x}^2+\xi^2+2 {\bar x}) (1-{\bar x})\over {(1-\xi^2)({\bar x}^2-\xi^2)}}\Big ]. 
\e
So we see that in contrast to the forward case, the effect of quark mass is not suppressed in the matrix element of the transverse component of the bilocal current. It is clear that such a helicity flip contribution is zero in the forward limit because the above expression is proportional to $\xi$. So, this effect is a particularity of the off-forward matrix element only.

An interesting study using this approach will  be to investigate the Wandzura-Wilczek relation \cite{ww} for the off-forward  matrix elements of the transverse vector and axial vector operators in perturbation theory,  for dressed quark states. We plan to undertake such studies in the near future.

To summarize, in this work, we have calculated the off-forward matrix element of the light-cone bilocal vector operator for a dressed quark state in perturbation theory. We have restricted ourselves in the kinematical region $\xi < {\bar x} <1$. The contribution from the overlap of three and one particle wave functions is absent in this case. We have obtained the corresponding splitting functions directly. The end point singularity is canceled by the contribution from the normalization condition of the state. We have shown that the generalized parton distributions $E_q$ and $E_g$ are suppressed in perturbation theory. Furthermore, we have verified the helicity sum sule in perturbation theory for a dressed quark state. The terms linear in quark mass cause helicity flip. However, such terms are suppressed in the matrix element of the plus component. We have calculated the helicity flip part of the matrix element of the transverse component of the same operator and explicitly shown that quark mass effects are not suppressed. We point out that it is a feature of the off-forward case only and this term is absent in the forward limit.

We would like to thank A. Harindranath and M. V. Polyakov for helpful discussions.


\begin{thebibliography}{99}

\bibitem{dieter} D. M\"uller, D. Robaschik, B. Geyer, F. M. Dittes, J.
Horejsi, Fortsch. Phys. {\bf 42}, 101 (1994).

\bibitem{ji} X. Ji, Phys. Rev. Lett. {\bf 78}, 610 (1997); Phys. Rev. {\bf D55}, 7114 (1997).

\bibitem{rad1} A. V. Radyushkin, Phys. Rev. {\bf D56}, 5524 (1997).

\bibitem{rad3} A. V. Radyushkin, Phys. Lett. {\bf B 385}, 333 (1996).

\bibitem{collins} J. C. Collins, L. Frankfurt, M. Strikman, Phys. Rev. {\bf
D 56}, 2982 (1997). 

\bibitem{rev} X. Ji, J. Phys. {\bf G 24}, 1181 (1998); A. V. Radyushkin,
hep-ph/0101225, published in "At the Frontier of Particle Physics/Handbook
of QCD", ed. M. Shifman (World Scientific, Singapore, 2001); K. Goeke, M.
V. Polyakov, M. Vanderhaeghen, Prog. Part. Nucl. Phys. {\bf 47}, 401 (2001). 

\bibitem{brod} S. J. Brodsky, M. Diehl, D. S. Hwang, Nucl. Phys. {\bf B 596},
99 (2001).

\bibitem{kroll} M. Diehl, T. Feldmann, R. Jakob, P. Kroll, Nucl. Phys. {\bf
B 596}, 33 (2001).

\bibitem{miller} M. Diehl, T. Feldmann, R. Jakob, P. Kroll, Eur. Phys. J {\bf C8}, 409 (1999);
B. C. Tiburzi and G. A. Miller, Phys. Rev. {\bf C 64},
065204 (2001); A. Mukherjee, I. V. Musatov, H. C. Pauli, A. V. Radyushkin, hep-ph/0205315. 

\bibitem{hari} A. Harindranath, R. Kundu, W. M. Zhang, Phys. Rev {\bf D 59},
094013; A. Harindranath, R. Kundu, A. Mukherjee and J. P. Vary, Phys. Lett.
{\bf B 417}, 361 (1997): Phys. Rev. {\bf D 58}, 114022 (1998): A.
Harindranath, A. Mukherjee, R. Ratabole, Phys. Lett. {\bf B 476}, 471
(2000); Phys. Rev. {\bf D 63}, 045006 (2001); A. Mukherjee, Phys. Lett. {\bf
B 517}, 109 (2001); A. Mukherjee and D. Chakrabarti, Phys. Lett. {\bf B
506}, 283 (2001).


\bibitem{oam}A. Harindranath and R. Kundu, Phys. Rev. {\bf D 59}, 116013 (1999).  

\bibitem{equal} X. Ji, J. Tang and P. Hoodbhoy, Phys. Rev. Lett. {\bf 76}, 740 (1996).

\bibitem{matrix} A. Harindranath and W. M. Zhang, Phys. Lett. {\bf B 390},
359 (1997).

\bibitem{ww} A. V. Belitsky and D. M\"uller, Nucl. Phys. {\bf B589}, 611
(2000); N. Kivel, M. V. Polyakov, A. Sch\"afer and O. V. Teryaev, Phys.
Lett. {\bf B497},73 (2001).
\end{thebibliography}
\end{document}